\documentclass[aps,prl,twocolumn,superscriptaddress,nofootinbib]{revtex4-1}

\usepackage{graphicx}
\usepackage{subfigure}
\usepackage{natbib}
\usepackage{amsmath,amssymb,amsfonts} 
\usepackage[usenames,dvipsnames]{xcolor}

\def \d {\mathrm{d}}

\renewcommand{\vec}[1]{\boldsymbol{\mathbf{#1}}} 
\renewcommand{\vec}[1]{\boldsymbol{\mathbf{#1}}}

\usepackage{verbatim}

\begin{document}


\title{Enhancement of the non-resonant streaming instability by particle collisions}



\author{A.~Marret}\affiliation{Sorbonne Universit\'e, Observatoire de Paris, Universit\'e PSL, CNRS, LERMA, F-75005, Paris, France}\affiliation{Sorbonne Universit\'e, Ecole Polytechnique, CNRS, Observatoire de Paris, LPP, F-75005, Paris, France}\affiliation{LULI, CNRS, Ecole Polytechnique, Sorbonne Universit\'e, CEA, Institut Polytechnique de Paris, F-91128 Palaiseau Cedex, France}

\author{A.~Ciardi}\affiliation{Sorbonne Universit\'e, Observatoire de Paris, Universit\'e PSL, CNRS, LERMA, F-75005, Paris, France}

\author{R.~Smets}\affiliation{Sorbonne Universit\'e, Ecole Polytechnique, CNRS, Observatoire de Paris, LPP, F-75005, Paris, France}

\author{J.~Fuchs}\affiliation{LULI, CNRS, Ecole Polytechnique, Sorbonne Universit\'e, CEA, Institut Polytechnique de Paris, F-91128 Palaiseau Cedex, France}

\author{L.~Nicolas}\affiliation{Sorbonne Universit\'e, Observatoire de Paris, Universit\'e PSL, CNRS, LERMA, F-75005, Paris, France}\affiliation{Sorbonne Universit\'e, Ecole Polytechnique, CNRS, Observatoire de Paris, LPP, F-75005, Paris, France}


\begin{abstract} 

Streaming cosmic rays can power the exponential growth of a seed magnetic field by exciting a non-resonant instability that feeds on their bulk kinetic energy. By generating the necessary turbulent magnetic field, it is thought to play a key role in the confinement and acceleration of cosmic rays at shocks. In this work we present hybrid-Particle-In-Cell simulations of the non-resonant mode including Monte Carlo collisions, and investigate the interplay between the pressure anisotropies produced by the instability and particle collisions in the background plasma. Simulations of poorly ionized plasmas confirm the rapid damping of the instability by proton-neutral collisions predicted by linear fluid theory calculations. In contrast we find that Coulomb collisions in fully ionized plasmas do not oppose the growth of the magnetic field, but under certain conditions suppress the pressure anisotropies and actually enhance the magnetic field amplification.

\end{abstract}
\maketitle

Ion streaming instabilities can develop when a population of energetic ions, such as cosmic rays, drifts at super-Alfvénic speeds in a background plasma permeated by a magnetic field \cite{kulsrudEffectWaveParticleInteractions1969,garyElectromagneticIonBeam1984,winskeDiffuseIonsProduced1984,bellTurbulentAmplificationMagnetic2004,amatoKineticApproachCosmicrayinduced2009}. The collective drifting motion drives the exponential growth of electromagnetic perturbations in many astrophysical \cite{volkMagneticFieldAmplification2005,amatoOriginGalacticCosmic2014,cuiYoungSupernovaRemnant2016}, space \cite{garyElectromagneticIonIon1991} and laboratory plasmas \cite{heuerObservationsFieldalignedIon2018}. Depending on the plasma conditions, three different modes of the instability exist \cite{garyElectromagneticIonBeam1984}: two of them rely on resonant particle-wave interactions, while the third mode is instead non-resonant (NR) and its importance was recognized in early work on the Earth's ion-foreshock \cite{sentmanInstabilitiesLowFrequency1981, onsagerInteractionFinitelengthIon1991,akimotoNonlinearEvolutionElectromagnetic1993}. More recently, the NR mode has become central to the diffusive shock acceleration of cosmic rays in supernovae remnants shocks, where it is thought to be able to amplify the magnetic field sufficiently to allow the confinement and acceleration of cosmic rays up to PeV energies \cite{bellTurbulentAmplificationMagnetic2004,amatoKineticApproachCosmicrayinduced2009,bellCosmicRayAcceleration2013}.
Studying the NR mode in the laboratory is also potentially within the reach of laser experiments with externally applied magnetic fields of tens of Tesla \cite{albertazziProductionLargeVolume2013}. 
Numerically, the instability has been extensively studied using modified magneto-hydrodynamics (MHD) \cite{bellTurbulentAmplificationMagnetic2004,zirakashviliModelingBellNonresonant2008}, hybrid-Particle-In-Cell (PIC ions and massless fluid electrons) \cite{winskeDiffuseIonsProduced1984,akimotoNonlinearEvolutionElectromagnetic1993,haggertyDHybridRHybridParticleincell2019,marretGrowthThermallyModified2021}, full-PIC \cite{riquelmeNonlinearStudyBell2009, ohiraTwodimensionalParticleincellSimulations2009, crumleyKineticSimulationsMildly2019} and MHD-PIC \cite{baiMagnetohydrodynamicParticleinCellMethodCoupling2015,vanmarleMagneticFieldAmplification2018, mignoneParticleModulePLUTO2018} simulations.

The MHD-PIC method has received growing attention as it combines the kinetic treatment of the cosmic rays while retaining the advantage of modelling the background plasma as a magnetofluid, over large spatial and temporal scales. Neglecting kinetic effects in the background plasma however is not always justified. For example, in the hot plasmas of superbubbles or in the intergalactic medium, the background's ions thermal Larmor gyro-radius can become comparable to or larger than the unstable wavelengths. Under these conditions the growth of the NR instability can be significantly reduced \cite{revilleEnvironmentalLimitsNonresonant2008,zweibelEnvironmentsMagneticField2010,marretGrowthThermallyModified2021}. In addition, collisionless hybrid-PIC simulations in cold plasmas have shown the development of significant ion pressure anisotropies in the background plasma \cite{marretGrowthThermallyModified2021}, suggesting that the assumption of an isotropic scalar pressure, often employed in fluid models, may not be sufficient. Pressure anisotropies can be suppressed by particle collisions or by micro-instabilities such as the mirror and ion-cyclotron mode, among other isotropization mechanisms. While ion-neutral collisions have been shown to damp the NR mode \cite{revilleCosmicRayCurrentdriven2007}, no studies have considered the effects of ion Coulomb collisions.

In this Letter we investigate the interplay between pressure anisotropies, micro-instabilities and particle collisions on the growth and saturation of the NR mode using hybrid-PIC simulations with Monte Carlo Collisions (MCC). We show that depending on the initial plasma-$\beta$, the magnetic field amplification driven by the NR mode can generate large pressure anisotropies. Because of the electromagnetic wave helical structure, these anisotropies give rise to pressure gradients that oppose the growth of the instability. We confirm that in poorly ionized plasmas ion-neutral elastic collisions damp the instability, as predicted by linear theory calculations. In fully ionized plasmas, we show that ion Coulomb collisions spanning over four orders of magnitude in collisions frequency do not hamper the instability, and unexpectedly enhance its growth by suppressing the self-generated pressure anisotropies. This leads to faster growth rates and 
to a larger amplification (up to $\approx 27\%$) of the magnetic field energy density with respect to the collisionless case.

The instability can be understood by considering a modified MHD model where a population of main protons and electrons (noted with the subscripts ``$m$'' and ``$e$'' respectively) behaves as a single, charged and incompressible background fluid, which supports a current compensating that of a less dense cosmic ray population (protons, noted with the subscript ``$cr$'') drifting with a velocity $u_{cr}$ parallel to an ambient magnetic field $B_0$ \cite{bellTurbulentAmplificationMagnetic2004}. 
Assuming purely growing modes, the instability may then be described \cite{marretGrowthThermallyModified2021} by the simplified linearized momentum conservation and Maxwell-Faraday's equations
\begin{align} 
\label{eq:momentum_heur}
\rho_0\dfrac{\partial\vec u_1}{\partial t} &= - \vec j_{cr}\times\vec B_1 \\
\label{eq:mag_heur}
\dfrac{\partial\vec B_1}{\partial t}&=(\vec B_0\cdot\vec\nabla)\vec u_1
\end{align}
where $\vec j_{cr}=en_{cr}\vec u_{cr}$ is the current carried by the cosmic rays, $e$ is the elementary charge, $\vec u$ is the fluid velocity and $\rho$ is the background plasma mass density. The subscripts ``0'' and ``1'' refer to the initial and perturbed quantities. The cosmic rays current is taken to be constant and the magnetic tension is neglected. 
\begin{figure}
	\includegraphics[width=\columnwidth]{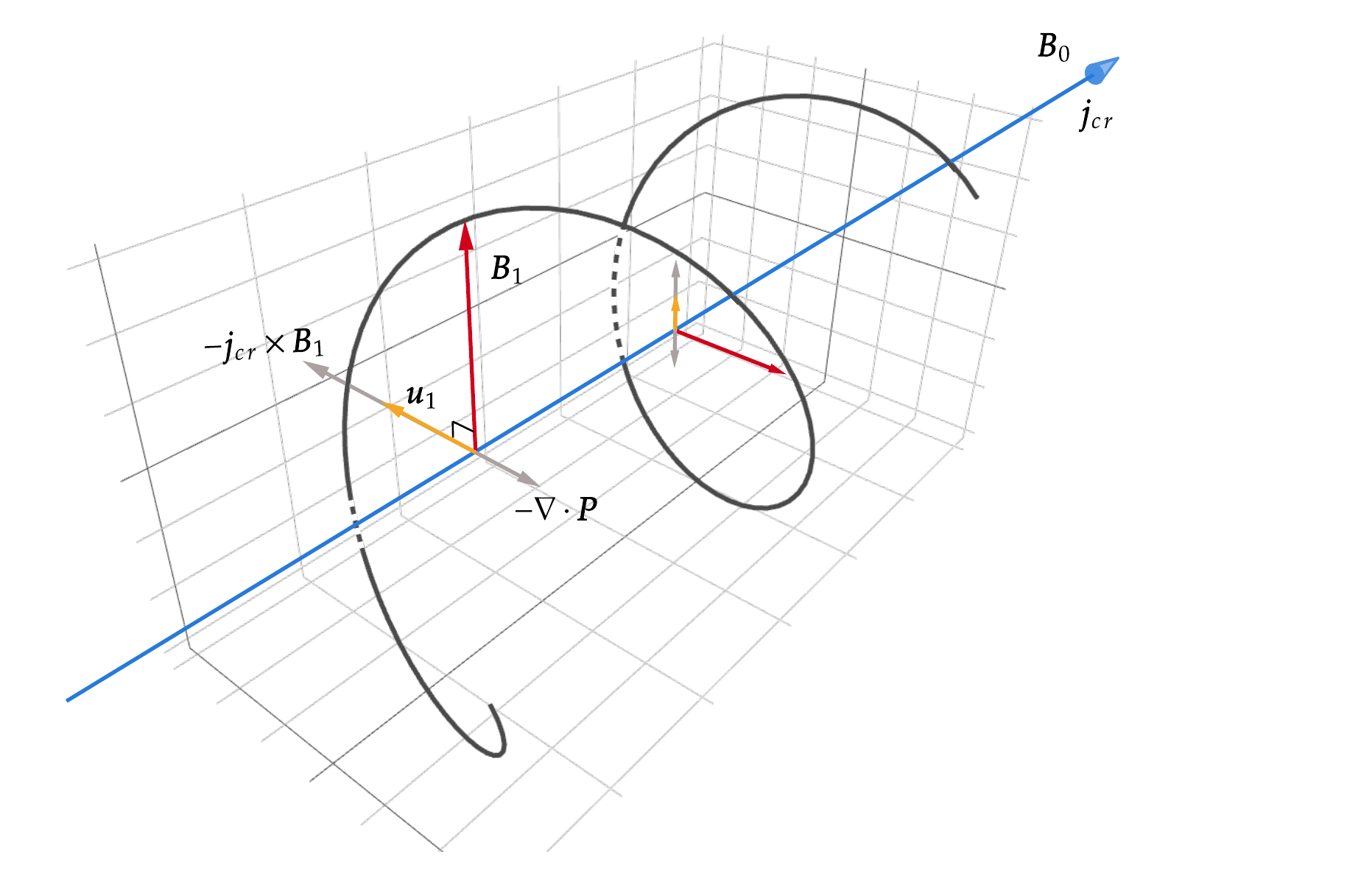}
    \caption{Schematic of the NR instability. The initial magnetic field $B_0$ is parallel to the cosmic rays current $j_{cr}$. The black solid line illustrates the magnetic field spatial structure for a right-hand polarized backward propagating electromagnetic wave. The instability is driven by the $-\vec j_{cr}\times \vec B_1$ force, and is opposed by the background pressure gradients force.}
    \label{fig:figure_1}
\end{figure}
The NR mode is driven by the cosmic ray current through the action of the magnetic force, $-\vec j_{cr}\times\vec B_1$, which produces fluid velocity fluctuations, $\vec{u}_1$, in the background plasma. The induced electric field, $-\vec u_1\times\vec B_0$, feeds back and enhances the initial magnetic field perturbation $B_1$ via Faraday's law, promoting its exponential growth. A schematic of the instability is shown in Fig. \ref{fig:figure_1}. 
Considering a right-hand circularly polarized magnetic field perturbations of the form $B_1e^{i(kx-\omega t)}$, where the angular frequency $\omega=\omega_r+i\gamma$, $\omega_r\ll\gamma$ is taken to be positive, the growth rate for negative $k$ is given by $\gamma_{\text{cold}}(k)=(\frac{n_{cr}}{n_m}\Omega_0u_{cr}|k|)^{1/2}$, where $u_{cr}$ is the cosmic rays velocity along $B_0$.
Although the background fluid pressure does not appear in the linear calculation for transverse electromagnetic fluctuations, we will show that non-linear pressure gradients effects can nonetheless modify the growth of the NR mode.

We present here the results of 1D and 2D simulations performed with the hybrid-PIC-MCC code HECKLE \cite{smetsr.HeckleHttpsGithub2011}. Details on the simulation parameters, normalizations, and the implementation of Monte Carlo collisions are given in the Supplemental Material which includes Refs. \cite{borisAccelerationCalculationScalar1970,holcombGrowthSaturationGyroresonant2019,takizukaBinaryCollisionModel1977,nanbuProbabilityTheoryElectronmolecule2000,krsticAtomicPlasmamaterialInteraction1999,recchiaGrammageCosmicRays2021,hunanaIntroductoryGuideFluid2019,trubnikovParticleInteractionsFully1965,garyProtonTemperatureAnisotropy1997}. We consider a population of cold main protons and electrons, traversed by a population of super-Alfv\'enic cosmic rays (protons) with a density $n_{cr}=0.01\ n_m$. The initial drift velocity, $u_{cr}= 50\ v_{A0}$, is oriented parallel to the initial magnetic field $\vec B_0=B_0\vec e_x$. nless stated otherwise, the initial plasma-$\beta$ is $\beta_0=P_0/W_{B0}=2$ where $P_0$ is the initial, isotropic main protons pressure and $W_{B0}=B_0^2/2\mu_0$. The plasma and field quantities are initially homogeneous, and periodic boundary conditions were used in all directions.

\begin{figure}
	\includegraphics[width=\columnwidth,trim=0 0cm 0 0,clip]{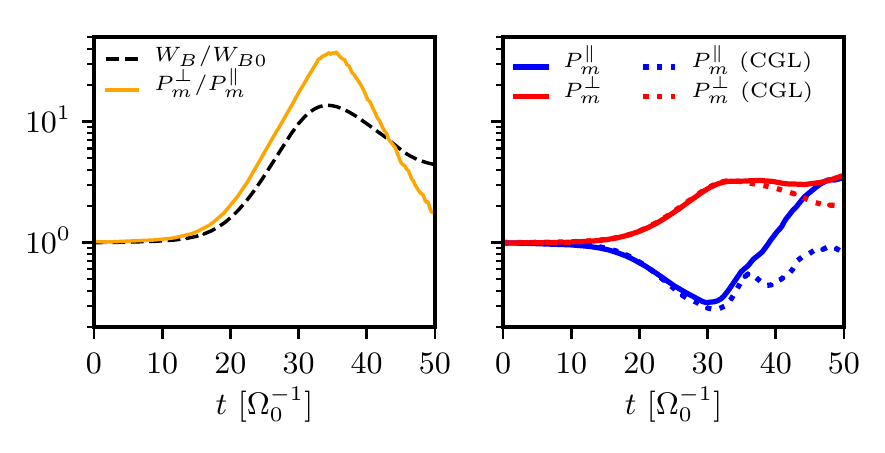}
    \caption{\textit{Left panel:} evolution of the spatially averaged magnetic field energy density $W_B$ (dashed black line) normalized to its initial value $W_{B0}=0.5\ l_0^{-3}m_pv_{A0}^2$ and the spatially averaged local ratio of pressure $P_m^\perp/P_m^{\smash[b]{\parallel}}$. Data is from a 1D collisionless simulation. \textit{Right panel}: evolution of the spatial average of the $P_m^\perp$ and $P_m^{\smash[b]{\parallel}}$ components of the main protons pressure tensor calculated from the local macroparticles distribution (red and blue solid lines). The dotted lines show the CGL predicted pressures $P_m^\perp/P_0=\rho B/\rho_0B_0$ and $P_m^{\smash[b]{\parallel}}/P_0=\rho^3B_0^2/\rho_0^3B^2$, calculated using the density and magnetic field from the simulations.
     }
    \label{fig:figure_2}
\end{figure}
In order to highlight the correlation between the growth of the NR mode and the generation of pressure anisotropies, we show the evolution of the magnetic field energy density $W_B=B^2/2\mu_0$ in the left panel of Fig. \ref{fig:figure_2} together with the ratio $P_m^\perp/P_m^{\smash[b]{\parallel}}$, corresponding to the ratio of the components of the main protons pressure tensor perpendicular and parallel to the total magnetic field.
The exponential growth of the magnetic perturbations between $t=15$ and $t= 34\ \Omega_0^{-1}$ is associated with important anisotropies, with a maximum spatially averaged value $P_m^\perp/P_m^{\smash[b]{\parallel}} = 36.7$. At saturation, the magnetic field energy density can be predicted by considering energy exchange rates obtained within quasi-linear theory \cite{winskeDiffuseIonsProduced1984}, which yield that the rate of energy gained by the magnetic field is half of the rate of loss of the cosmic rays drift kinetic energy. Extrapolating this result to saturation and supposing that the initial cosmic rays drift kinetic energy density $W_{cr}=n_{cr}m_pu_{cr}^2/2$ is entirely depleted at saturation, one obtains $W_{B,\text{sat}} = W_{cr}/2$, which is close to the simulations results $W_{B,\text{sat}} =0.55\ W_{cr}$ obtained by averaging the magnetic field energy density over space.

The development of pressure anisotropies can be described within the adiabatic CGL theory \cite{chewBoltzmannEquationOnefluid1956}, which may be interpreted as the conservation of the first and second adiabatic invariants in an amplified magnetic field \cite{kulsrudMHDDescriptionPlasma1983}. Assuming cold electrons, neglecting the Hall effect, heat fluxes and non-gyrotropic pressure components, 
the CGL equations for the main protons are $\d_t(P_m^{\smash[b]{\parallel}} B^2/\rho^3)=0$ and $\d_t(P_m^\perp /\rho B)=0$ where $\d_t=\partial_t+\vec u_m\cdot\vec\nabla$ denotes the material derivative. 
The advective term may be neglected by integrating over the periodic simulation domain and assuming small density fluctuations, which is a well verified in the simulations, allowing to evaluate the pressure components directly.
Fig. \ref{fig:figure_2} compares the evolution of the pressure components $P_m^\perp$ and $P_m^{\smash[b]{\parallel}}$ predicted from the CGL model and calculated using the simulated density and magnetic field, with the corresponding components of the full pressure tensor obtained from the macroparticle's velocity distribution. The pressure anisotropies driven by the NR mode evolve according to the the CGL equations up to saturation ($\sim 35\ \Omega_0^{-1}$), when the subsequent development of important non-gyrotropic pressure components invalidates the CGL approximations.

\begin{figure}
\centering    
\includegraphics[width=\columnwidth]{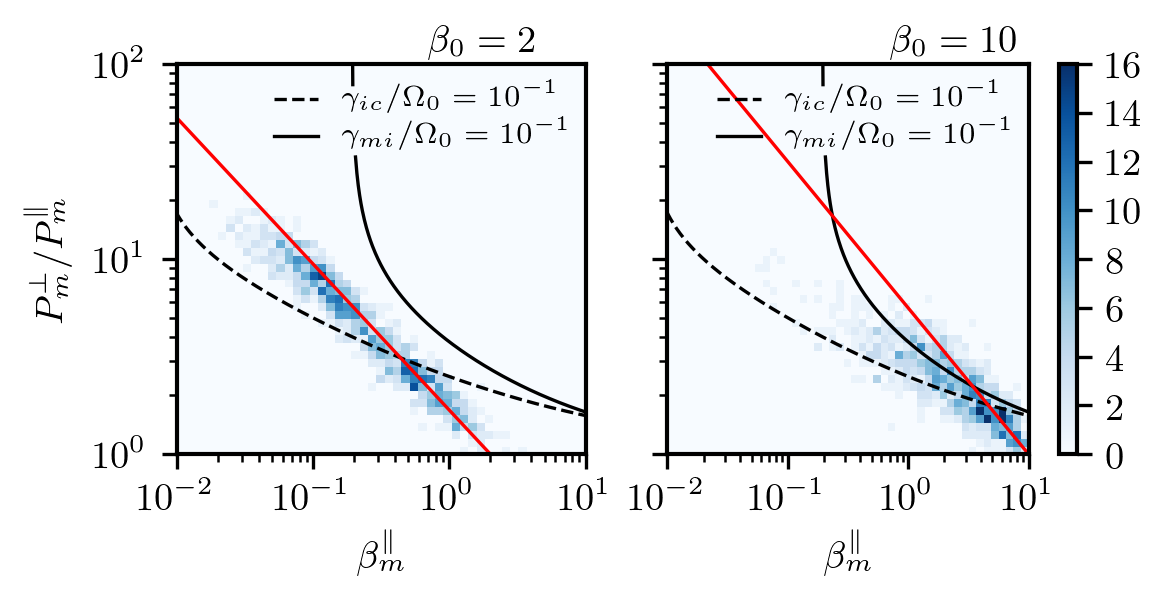}
\caption{Distribution (cell count) of the ratio $P_m^\perp/P_m^{\smash[b]{\parallel}}$ as a function of $\beta_m^\parallel=P_m^{\smash[b]{\parallel}}/W_B$, obtained in 1D simulations without collisions during the exponential phase of growth. \textit{Left panel:} $\beta_0=2$ at $t=25\ \Omega_0^{-1}$. \textit{Right panel:} $\beta_0=10$ at $t=35\ \Omega_0^{-1}$. We increased the plasma-$\beta$ by increasing the main protons initial temperature, which also introduces finite Larmor-radius effects that reduce the instability growth rate \cite{zweibelEnvironmentsMagneticField2010,marretGrowthThermallyModified2021} hence the difference in times in the two panels. The solid red line corresponds to the anisotropy expected from incompressible CGL theory $P_m^\perp/P_m^{\smash[b]{\parallel}}=(\beta_m^\parallel/\beta_0)^{-3/4}$. The solid and dashed black lines indicate the thresholds for the mirror $\gamma_{mi}=10^{-1}\ \Omega_0$ and ion-cyclotron $\gamma_{ci}=10^{-1}\ \Omega_0$ modes respectively, obtained from linear kinetic theory assuming a homogeneous plasma with bi-Maxwellian populations \cite{hellingerSolarWindProton2006}, for a growth rate comparable to that of the collisionless NR mode ($\gamma_0=0.15\ \Omega_0$). The micro-instabilities growth rates increase with larger $P_m^\perp/P_m^{\smash[b]{\parallel}}$, and are stabilized for smaller $\beta_m^\parallel$.} 
\label{fig:figure_6}
\end{figure}

Pressure anisotropies $P_m^\perp/P_m^{\smash[b]{\parallel}}>1$ are known to be responsible for the growth of the ion-cyclotron and mirror micro-instabilities \cite{garyProtonTemperatureAnisotropy1976}.
To understand their role, we display in Fig. \ref{fig:figure_6} the distribution (cell count) of the ratio $P_m^\perp/P_m^{\smash[b]{\parallel}}$ as a function of $\beta_m^\parallel=P_m^{\smash[b]{\parallel}}/W_B$, for two collisionless simulations with $\beta_0=2$ and $\beta_0=10$. The incompressible CGL theory discussed earlier implies that the anisotropies driven by the NR mode should follow a power law dependence given by  $P_m^\perp/P_m^{\smash[b]{\parallel}}=(\beta_m^\parallel/\beta_0)^{-3/4}$ (indicated with a red line).  Fig. \ref{fig:figure_6} also shows the threshold anisotropy for the ion-cyclotron and mirror instabilities, for a growth rate comparable to that of the NR mode.
We find that in the simulations with $\beta_0=2$ (Fig. \ref{fig:figure_6} left panel), the power law behaviour is indeed well recovered, indicating that the anisotropies are not constrained by these micro-instabilities. In particular, for the mirror mode the growth rate, $\gamma_{mi}\sim 10^{-2}\ \Omega_0$,  always remains below that of the NR mode, $\gamma_0= 0.15\ \Omega_0$.

A different picture however emerges for the case $\beta_0=10$, shown in the right panel of Fig. \ref{fig:figure_6}, where for $P_m^\perp/P_m^{\smash[b]{\parallel}} \gtrsim 2$ the growth of the mirror mode competes with the NR mode ($\gamma_{mi}\sim\gamma$) and keeps the pressure anisotropies at the threshold values, below the expected power law. We note that although in both cases the ion-cyclotron instability has nominally a growth rate larger than the mirror mode, and comparable or larger to that of the NR mode  for $P_m^\perp/P_m^{\smash[b]{\parallel}} \gtrsim 2-3$, it does not appear to be limiting the anisotropy; a similar behaviour is also observed in the solar wind \cite{hellingerSolarWindProton2006}. In our case it may be due to the spatial inhomogeneity of the magnetic field amplification, which renders the cyclotron resonance position dependent and impairs the growth of the ion-cyclotron mode \cite{southwoodMirrorInstabilityPhysical1993}.

\begin{figure}
	\includegraphics[width=\columnwidth]{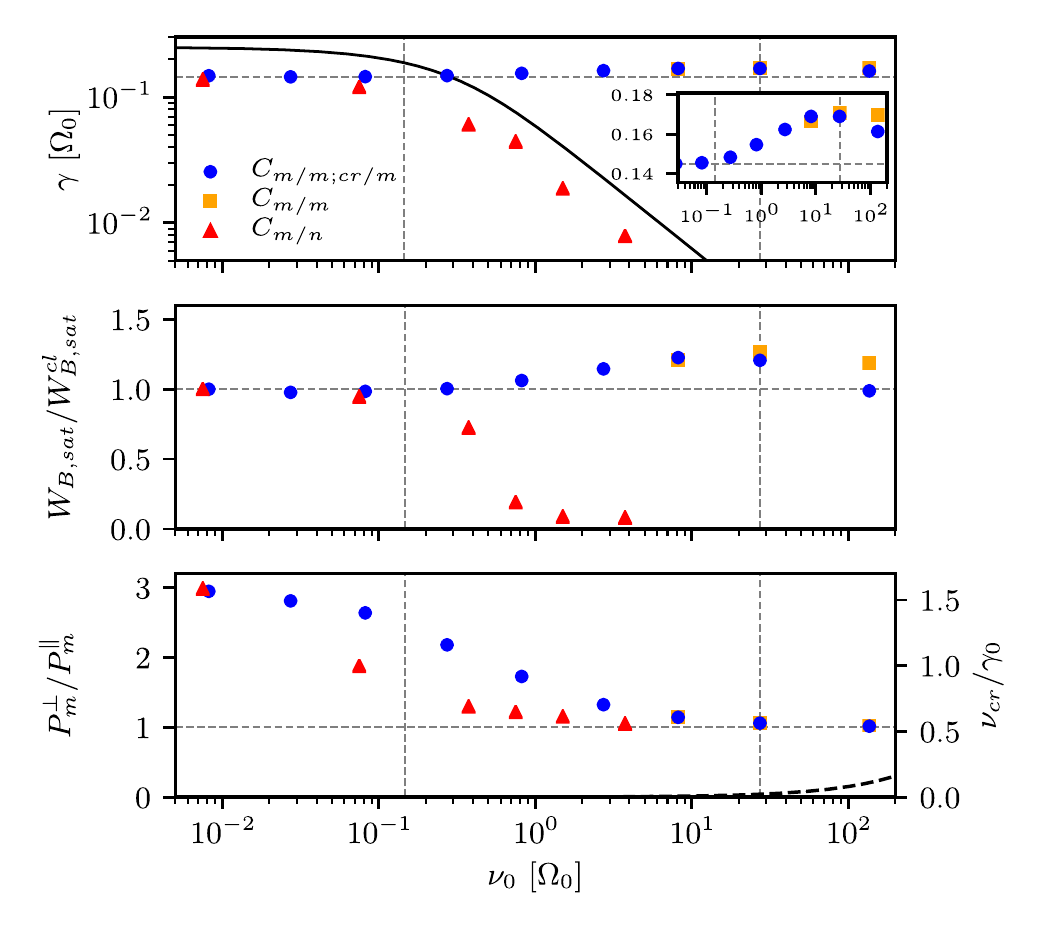}
    \caption{\textit{Upper panel}: Magnetic field growth rate $\gamma(\nu_0)$, in 1D runs with Coulomb collisions between all protons populations ($C_{m/m;cr/m}$, blue dots) and with Coulomb collisions between main protons only ($C_{m/m}$, orange squares) where $\nu_0=e^4n_m\ln{\Lambda}/4\pi m_p^2\epsilon_0^2 v_{T0}^3$ is the collision frequency among the main proton with $\epsilon_0$ the vacuum permittivity, $\ln{\Lambda}$ is the Coulomb logarithm, and $v_{T0}=(k_B T_0/m_p)^{1/2}$ the thermal velocity with $k_B$ the Boltzmann constant. The inset highlights the increase in growth rate with Coulomb collision frequency. The red triangles indicate simulations with main proton-neutral collisions ($C_{m/n}$), where $\nu_{0}=n_{n}\sigma_nv_{T0}$ is the main proton-neutral collision frequency, $n_n$ is the neutral density and $\sigma_{n}$ is the neutral collision cross-section. The solid black line corresponds to the theoretical growth rate $\gamma_{in}$ considering $k=k_{\max}/4$. The magnetic field growth rate in the collisionless case $\gamma_0=0.15\ \Omega_0$ is indicated with the first vertical dashed line. The maximum growth rate for simulations with Coulomb collisions is indicated with the second vertical dashed line at $\nu_0=27\ \Omega_0$. \textit{Middle panel}: Magnetic field energy density $W_{B,\text{sat}}=B_{\text{sat}}^2/2\mu_0$ at saturation, normalized to the value in collisionless simulations $W_{B,\text{sat}}^{cl}=6.84\ l_0^{-3}m_pv_{A0}^2$. \textit{Lower panel}: Mean value of the ratio $P_m^\perp/P_m^{\smash[b]{\parallel}}$ averaged over the exponential phase of growth. The dashed black line (bottom right) corresponds to the initial cosmic ray-main proton collision frequency $\nu_{cr}$, normalized to 
    $\gamma_0$.}
    \label{fig:figure_3}
\end{figure}

In a collisional plasma, pressure anisotropies may be mitigated if collisions are sufficiently frequent to redistribute the energy in all directions of space. 
Here we investigate two cases: a fully ionized background plasma such that ion Coulomb collisions are dominant, and a poorly ionized background where collisions with a population of neutrals are dominant.
Fig. \ref{fig:figure_3} upper panel presents the magnetic field growth rate $\gamma$ as a function of the reference collision frequency $\nu_0$. 
The growth rate in the case of a weakly ionized plasma is given by $\gamma_{in}(k)=-\frac{\nu_{0}}{2}+\frac{1}{2}(\nu_{0}^2+4\Omega_0\frac{n_{cr}}{n_m}ku_{cr})^{1/2}
\label{eq:growth_neutral}$ \cite{revilleCosmicRayCurrentdriven2007}. This is plotted in Fig. \ref{fig:figure_3} (solid black line in upper panel) considering $k=k_{\max}/4=\frac{1}{4}\frac{n_{cr}}{n_m}\frac{u_{cr}}{v_{A0}^2}\Omega_0$, such that $\gamma_{in}(\nu_0=0)=\gamma_{\max}$, where $\gamma_{\max}=\frac{1}{2}\frac{n_{cr}}{n_m}\frac{u_{cr}}{v_{A0}}\Omega_0$ is the fastest growing mode in the collisionless case \cite{winskeDiffuseIonsProduced1984}. 
The growth rate dependency with collision frequency is well recovered in the simulations (red triangles). The magnetic field intensity in the simulations is integrated over the $k$ spectrum before measuring the growth rate, in order to reduce noise due to the dynamic nature of the range of unstable wavenumbers inherent to the NR mode \cite{bellTurbulentAmplificationMagnetic2004,marretGrowthThermallyModified2021}. This gives an overall smaller growth rate than if only the fastest growing mode was observed and is seen with the offset in the figure.

For simulations of a fully ionized collisional background (blue dots), we find that the growth rate is enhanced with respect to the collisionless case for $\nu_0 > \gamma_0$, where $\gamma_0=0.15\ \Omega_0$. The increase is maximum for a collision frequency $\nu_0=27\ \Omega_0$ two orders of magnitude larger than $\gamma_0$, yielding a growth rate $\gamma=0.17\ \Omega_0$. The saturated magnetic field energy density, $W_{B,\text{sat}}$ is displayed in Fig. \ref{fig:figure_3} middle panel, and shows an increase up to $W_{B,\text{sat}}=8.65\ l_0^{-3}m_pv_{A0}^2$ corresponding to $W_{B,\text{sat}}=1.27\ W_{B,\text{sat}}^{cl}$ with $W_{B,\text{sat}}^{cl}$ the saturated magnetic field energy density in the colisionless case. The most unstable wavenumber is found to be insensitive to the collision frequency, with unstable waves growing on scales of the order $\lambda\approx 4\pi k_{\max}^{-1}\approx 25\ l_0$ in agreement with the linear kinetic theory prediction for a negligible background plasma temperature \cite{winskeDiffuseIonsProduced1984}. Similar results are obtained in 2D simulations (see the Supplemental Material).

Because of the relatively large density of cosmic rays in the simulations, their collisions with the background protons become important for $\nu_0\gg 10\ \Omega_0$, and leads to a rapid reduction of their relative drift velocity and a lower magnetic field amplification, as seen in Fig. \ref{fig:figure_3} middle panel.
This was verified by performing simulations where cosmic rays collisions were artificially suppressed (orange squares in Fig. \ref{fig:figure_3}). In the case of completely collisionless cosmic rays, which is also more representative of the conditions found in space, the growth rate and magnetic field energy at saturation remain at the same level.

The lower panel of Fig. \ref{fig:figure_3} shows the ratio $P_m^\perp/P_m^{\smash[b]{\parallel}}$ as a function of the collision frequency $\nu_0$. The observed increase in the amplification of the magnetic field with $\nu_0$, corresponds to the gradual suppression of the pressure anisotropies for $\nu_0 \gtrsim 0.1\ \Omega_0 \sim \gamma_0$. 
This is consistent with an estimate comparing the anisotropic heating rate in the low plasma-$\beta$ limit to the collisional isotropization rate: $\nu_0 / \gamma >\kappa^{-1}$,
where $\kappa=3(2\pi^{1/2}A)^{-1}\{-3+(A+3)[\tan^{-1}(A^{1/2})/A^{1/2}]\}$ is a decreasing function of $A=P_m^{\perp}/P_m^{\smash[b]{\parallel}}-1$. Details on the derivation are given in the Supplemental Material. For our simulation parameters $\kappa^{-1}=7.3$, which agrees reasonably well with the range of collision frequencies, $\nu_0/\gamma>1$, for which pressure anisotropies are seen to be strongly reduced. Keeping anisotropies small, say $P_m^\perp/P_m^{\smash[b]{\parallel}}-1< 0.1$, requires collision frequencies $\nu_0/\gamma>10^2$, which is again consistent with the values obtained in the simulations.

\begin{figure}
\includegraphics[width=\columnwidth]{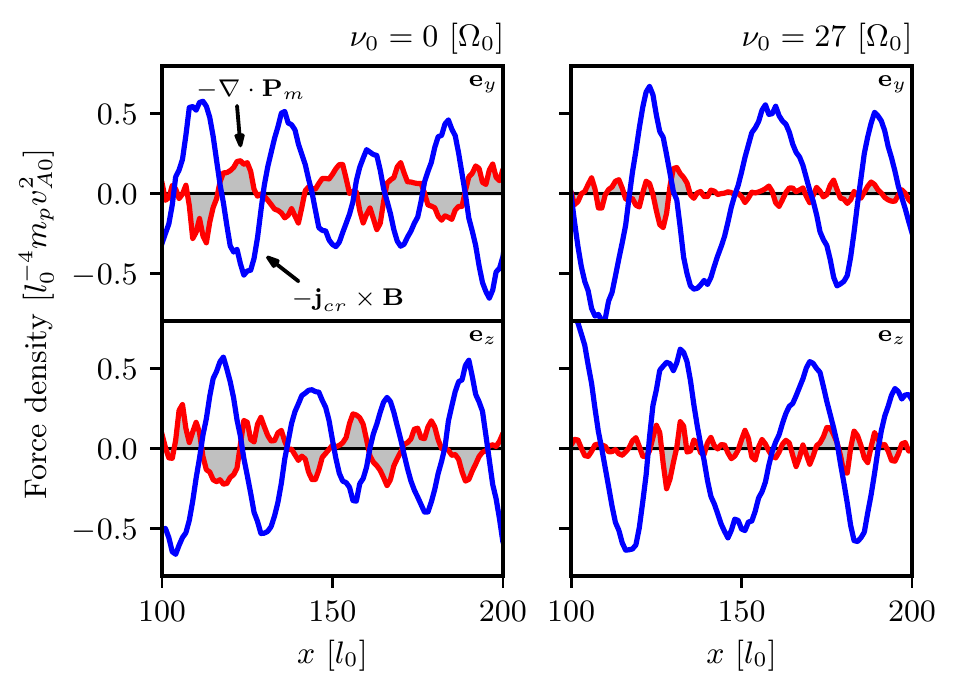}
\caption{Cosmic rays induced magnetic force $-\vec j_{cr}\times\vec B$ (blue solid line) and main protons pressure gradients $-\vec\nabla\cdot\vec P_m$ (red solid line) components in the plane perpendicular to $B_0$, as a function of space between $x=100$ and $x=200\ l_0$. Data is taken during the exponential growth phase at $t=30\ \Omega_0^{-1}$ in a 1D collisionless simulation (upper and lower left panels), and with Coulomb collisions at a frequency $\nu_0=27\ \Omega_0$ (upper and lower right panels). The grey regions highlight the pressure gradients contribution, convoluted with a Gaussian to reduce fluctuations at the mesh size scale $l_0$ in the figure.}
\label{fig:figure_4}
\end{figure}

The enhanced amplification of the magnetic field can be understood by considering the electromagnetic wave helical structure in space: the pressure anisotropies generate spatial gradients of the pressure tensor along $B_0$, which in turn affect the background plasma dynamics in the plane perpendicular to $B_0$ and oppose the magnetic force driving the NR mode. Suppressing these anisotropies promotes the growth of the magnetic field, however, we do not expect that a strong enhancement ($>100\%$) is possible since it would require the pressure gradients to overcome the cosmic rays magnetic force, which would prevent the growth of the instability altogether.
The competition between the magnetic and the pressure gradient forces is presented in Fig. \ref{fig:figure_4} for 1D collisionless and collisional simulations. It is clear that the force $-\vec j_{cr}\times\vec B$ is opposed by the pressure tensor gradients of the background protons $-\vec\nabla\cdot\vec P_m$, as illustrated in Fig. \ref{fig:figure_1}. We find $|\vec j_{cr}\times\vec B|/|\vec\nabla\cdot\vec P_m|\sim 3$ during the exponential phase of growth in the collisionless case, which leads to a less efficient acceleration of the background fluid and consequently to a proportionally smaller magnetic field amplification (Eq. \ref{eq:mag_heur}). We verified that collisional viscous forces are negligible in the simulations. 

The implications of the present study on the acceleration of cosmic rays at shocks remain to be explored. Recent large scale 3D MHD-PIC simulations \cite{vanmarleMagneticFieldAmplification2018} have highlighted the complexity of the shock structure and dynamic, as well as the importance of the back-reaction of accelerated cosmic rays on the shock itself. Extensions to the MHD-PIC model beyond isotropic pressure or hybrid-PIC simulations will be needed to assess the role of pressure anisotropies, as well as micro-instabilities \cite{leyStochasticIonAcceleration2019}, in this context.

\section{Acknowledgments}
\begin{acknowledgments}
The authors would like to thank Stefano Gabici, Alexandre Marcowith and Mark Sherlock, as well as the anonymous referees, for useful discussions and comments on this work. This work was performed using HPC resources from Grand Equipement National de Calcul Intensif- [Tr\`es Grand Centre de Calcul] (Grant 2019- [DARI A0060410819]), and was granted access to the HPC resources of MesoPSL financed by the Region Ile de France and the project EquipMeso (reference ANR-10-EQPX29-01) of the programme Investissements d'Avenir supervised by the Agence Nationale pour la Recherche. This work was partly done within the Plas@Par LABEX project and supported by grant 11-IDEX-0004-02 from Agence Nationale pour la Recherche (France), and by Grant ANR-17-CE30- 0026-Pinnacle from Agence Nationale de la Recherche and by the European Research Council (ERC) under the European Union’s Horizon 2020 research and innovation program (Grant Agreement No. 787539).
\end{acknowledgments}

\section*{Supplemental Material}
\section*{The HECKLE code and normalizations}
We use the hybrid-PIC code HECKLE \cite{smetsr.HeckleHttpsGithub2011}, which solves the Vlasov-Maxwell system using a predictor-corrector scheme for the electromagnetic field and a non-relativistic Boris pusher \cite{borisAccelerationCalculationScalar1970} for the particles. The main protons and cosmic rays protons are described as macro-particles following the PIC method, and the electrons are treated as a mass-less fluid. This hybrid approach is well suited to study the kinetic, non-linear evolution of systems at the ions temporal scale while avoiding prohibitive computational time and is expected to be sufficient to study the NR instability, as supported by quasi-linear theory which predicts that the heating and momentum exchange rates of the background electrons are much smaller than those of the ions \cite{winskeDiffuseIonsProduced1984}. In the simulations, the densities and magnetic field are normalized to the initial uniform values $n_0=n_m(t\!=\!0)$ and $B_0 = B(t\!=\!0)$. Times and lengths are normalized to the inverse of the initial proton cyclotron angular frequency $\Omega_0=eB_0/m_p$, where $e$ and $m_p$ are the elementary charge and proton mass respectively, and initial proton inertial length $l_0=c/\omega_{pm}$ where $c$ is the speed of light and $\omega_{pm}=(n_0e^2/\varepsilon_0 m_p)^{1/2}$ is the protons plasma frequency. Velocities are normalized to the Alfv\'en velocity $v_{A0}=B_0/(\mu_0n_0m_p)^{1/2}=l_0\Omega_0$. 

\section*{Numerical setup}
The cosmic rays population of density $n_{cr}/n_0=0.01$ is given an initial positive drift velocity in the background plasma reference frame,  $u_{cr}/v_{A0}=50$. 
The initial magnetic field $\vec B_0$ is oriented parallel to the cosmic rays drift velocity, aligned with the $\vec e_x$ direction. The initial temperature is $k_BT=m_pv_{A0}^2$ for all the populations, corresponding to $\beta_0=2$. We also ran a simulation with a larger main protons temperature $k_BT_m=5m_pv_{A0}^2$ such that $\beta_0=10$. The electrons population is taken to be isothermal. The precise shape of the cosmic rays distribution function is unimportant for the non-resonant mode (\cite{garyElectromagneticIonBeam1984,bellTurbulentAmplificationMagnetic2004}), owing its non-resonant nature. As such, the cosmic rays drifting population is modelled in the simulations using a cold Maxwellian distribution, without any loss of physical accuracy. This model cannot be used when studying the left-hand and right-hand resonant modes, where details of the cosmic rays streaming population are important in determining the growth of the electromagnetic perturbations (e.g. \cite{holcombGrowthSaturationGyroresonant2019}).

The condition of a null initial current is achieved by considering a drift velocity for the electrons population relative to the main protons, in the same direction as the cosmic rays. A different way of compensating the current would be to distinguish two electrons populations: one with the same density as the main protons, and an additional population with the same charge density as the cosmic rays and drifting alongside them.  It can be shown within the framework of kinetic theory \cite{amatoKineticApproachCosmicrayinduced2009} that the dispersion relation of the NR mode is only modified by a corrective term of the order $O(n_{cr}^2/n_m^2)$ depending on the choice to compensate the current. Hence the relatively large density ratio $n_{cr}/n_m=0.01$ in the simulations, larger than the typical ratio expected in astrophysical environments, is still relevant to the case of magnetic field amplification by the cosmic rays driven NR instability, regardless of the current compensation scenario.

The simulation domain has a length $L_x=1000$ $l_0$ where $l_0$ is the proton inertial length and discretized with 1000 cells for 1D simulations, extended to $L_y = 200$ $l_0$ in the 2D case with the same grid size as in the $x$ direction. The plasma and field quantities are initially homogeneous, and periodic boundary conditions are used in all directions. In 1D simulations we used 200 macroparticles per cells, 100 for each proton species. For 2D simulations, we used 600 macroparticles per cell, 500 for the background protons and 100 for the cosmic rays. Each species possesses a numerical weight, which is used to deposit the moments of the distribution function on the grid while taking into account for the different densities between background protons and cosmic rays.

The proton-proton Coulomb collisions are implemented numerically using a Monte Carlo method which solves the Landau collisions operator by randomly pairing macroparticles in each cells, and calculating at each time step the associated scattering angle and post-collision velocities \cite{takizukaBinaryCollisionModel1977}.

The proton-neutral collisions are implemented following a hard-sphere model \cite{nanbuProbabilityTheoryElectronmolecule2000} and adapted for the hybrid-PIC approach. We considered proton-Hydrogen elastic collisions, and a small ionisation fraction such that the neutrals density and temperature are supposed to remain constant and uniform. At each time step a collision between a macroparticle and an hydrogen atom may occur if the condition $r<n_n\sigma_{in}\Delta v\Delta t$ is verified, where $r$ is a random number between 0 and 1, $n_n$ is the neutral density, $\sigma_{in}$ is the collision cross-section, $\Delta v$ is the relative velocity and $\Delta t$ the time step. The cross-section collision energy dependency is obtained from Ref. \cite{krsticAtomicPlasmamaterialInteraction1999}. The initial background proton thermal energy was taken as $k_BT_m=m_pv_{A0}^2=1\ \text{eV}$. We note that for smaller energies the charge exchange collisions, where the ion takes an electron from the neutral particle, cannot be dissociated from the elastic collisions for proton-hydrogen interactions \cite{recchiaGrammageCosmicRays2021}. In such a case the friction force used in Ref. \cite{revilleEnvironmentalLimitsNonresonant2008} to derive the NR instability growth rate while considering ion-neutral collisions must be modified. The Coulomb and neutral collision frequencies, noted $\nu_0$, can be scaled with respect to the instability growth rate $\gamma$, allowing to probe the weakly ($\gamma>\nu_0$) and strongly ($\gamma<\nu_0$) collisional regimes of the NR mode. The electron resistivity is negligible for the range of Coulomb collision frequencies investigated, from $\nu_0=10^{-2}$ to $\nu_0=10^2\ \Omega_0$. Larger values were not investigated because of the prohibitively small numerical time steps required to resolve low energy collisions.

\section{Magnetic field amplification in 2D simulations}
\begin{figure}
	\includegraphics[width=\columnwidth]{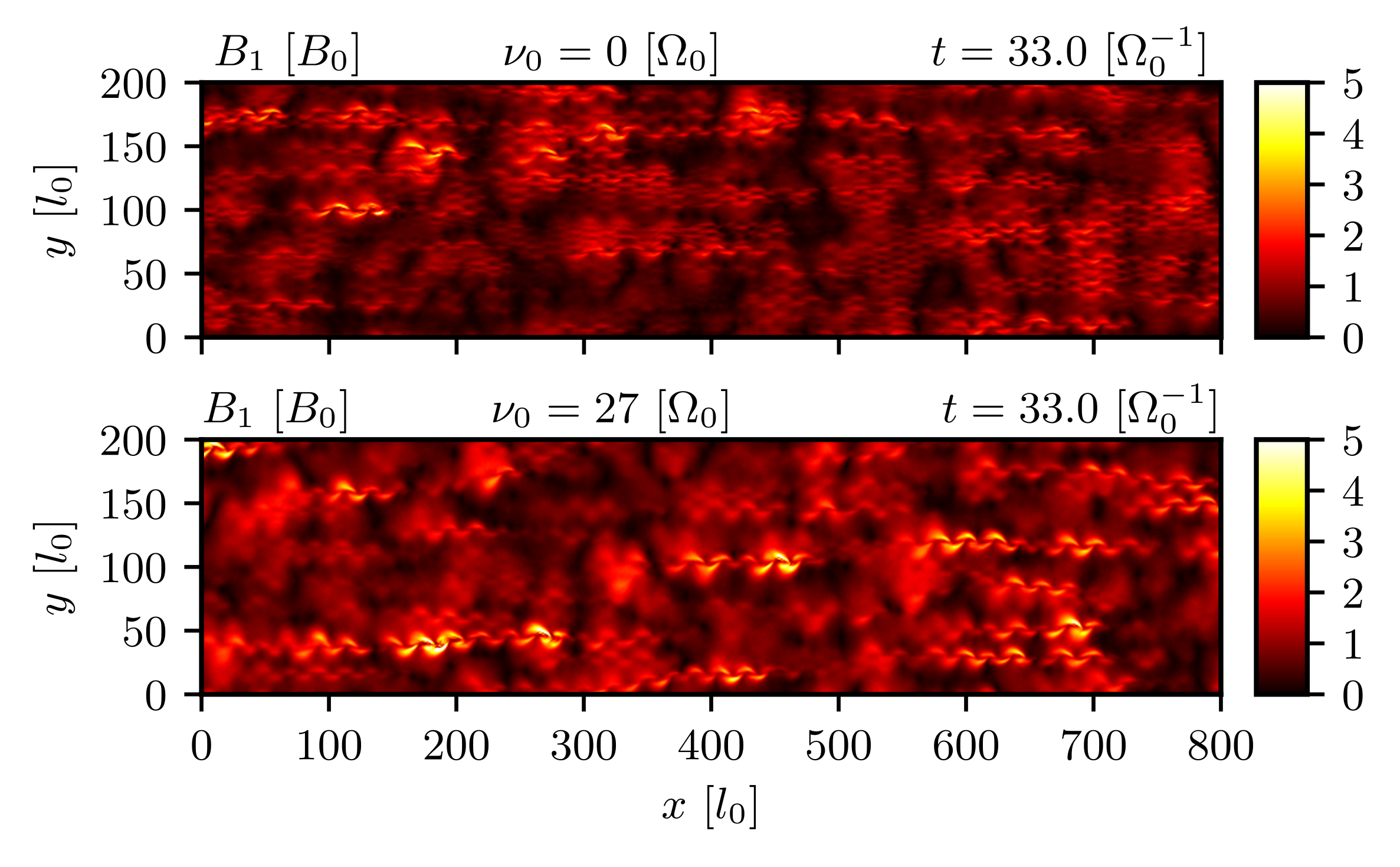}
    \caption{Map of the perturbed magnetic field intensity $B_1$, in units of $B_0$, during the exponential phase of growth in 2D simulations between $x=0$ and $x=800\ l_0$. \textit{Upper panel}: Collisionless simulation. \textit{Lower panel}: Including Coulomb collisions with a collision frequency $\nu_0 =27\ \Omega_0$.
    }
    \label{fig:figure_5}
\end{figure}
\begin{figure}
	\includegraphics[width=\columnwidth,trim=0 0cm 0 0.2cm,clip]{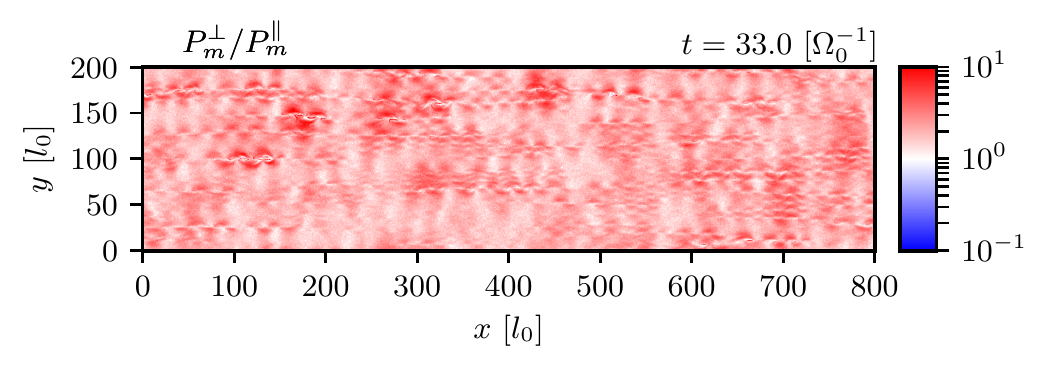}
    \caption{Map of the ratio $P_m^\perp/P_m^\parallel$ at the end of the exponential phase of growth in a 2D collisionless ($\nu_0= 0\ \Omega_0$) simulation between $x=0$ and $x=800\ l_0$.
     }
    \label{fig:figure_2_2}
\end{figure}
In 2D simulations, we find a growth rate of the magnetic field intensity averaged over space marginally larger ($\sim 1\%$) in the collisional case. The magnetic field is amplified similarly to 1D simulations, with an average magnetic field energy density ratio between the collisional and collisionless simulations of $W_{B,\text{sat}}=1.3\ W_{B,\text{sat}}^{cl}$ over the growth of the instability. This is illustrated in Fig. \ref{fig:figure_5}, which displays maps of the perturbed magnetic field intensity for 2D collisionless and collisional ($\nu_0=27\ \Omega_0$) simulations. The regions of magnetic field amplification are exactly correlated with the regions of large pressure anisotropies, as shown in Fig. \ref{fig:figure_2_2}, confirming that the CGL predictions describe well the observed anisotropies.

\section{Collisional isotropization and pressure gradients estimates}
One may estimate the Coulomb collision frequency required to mitigate the pressure anisotropies by comparing the NR mode anisotropic heating rate to the isotropization rate by collisions. The evolution of the anisotropy, $P_m^\perp-P_m^\parallel$, within CGL theory and including collisions can be expressed as
\begin{equation}
\dfrac{\partial }{\partial t}(P_m^\perp-P_m^\parallel) = \gamma (P_m^\perp+2P_m^\parallel) -\nu_0 \kappa P_0^{3/2} \dfrac{1}{\sqrt{\smash[b]{P_m^\parallel}}}
\label{eq:anisotropic_evolution}
\end{equation}
where the first term on the right hand side is the anisotropic heating rate due to the amplification of the magnetic field \cite{hunanaIntroductoryGuideFluid2019} by the NR mode, and the second term is the pressure isotropization rate due to Coulomb collisions \cite{trubnikovParticleInteractionsFully1965}. The magnetic field growth rate is defined as $\gamma\equiv(\d B/\d t)/B$, with $B=|\vec B|$; $\nu_0$ is the fundamental Coulomb collision frequency, and $\kappa=3(2\pi^{1/2}A)^{-1}\{-3+(A+3)\left[\tan^{-1}(A^{1/2})/A^{1/2}\right]\}$ is a decreasing function of $A=P^{\perp}_m/P^{\parallel}_m-1$. The above expression is valid in the incompressible limit, and we have also neglected the advection term. 
Inserting the incompressible CGL equations in Eq. \ref{eq:anisotropic_evolution}, and considering that for the NR instability $(B/B_0)^2\gg1$, one then obtains the level of collisionality necessary to stop the growth of the pressure anisotropies and reach steady state
\begin{equation}
\dfrac{\nu_0}{\gamma} = \dfrac{1}{\kappa}
\label{eq:nu_gamma_steady}
\end{equation}
while larger collisions frequencies, i.e. $\nu_0 / \gamma > \kappa^{-1}$, will start to strongly affect the development of pressure anisotropies. The function $\kappa$ requires to calculate the parameter $A$, which may be inferred from the saturated magnetic field energy density prediction $W_B\approx W_{cr}/2$ obtained from quasi-linear theory \cite{winskeDiffuseIonsProduced1984}, such that $A=(W_{cr}/2W_{B0})^{3/2}-1$.

The effect of pressure anisotropies can be estimated by comparing the pressure gradient force to the magnetic force driving the instability as $k_{\text{fast}}(P_m^\perp-P_m^{\smash[b]{\parallel}}) \gtrsim j_{cr}B$ with $k_{\text{fast}}=k_{\max}/2$ the fastest growing wavenumber \cite{bellTurbulentAmplificationMagnetic2004}. Inserting the incompressible CGL pressure equations
and considering $(B/B_0)^2\gg1$, 
the condition for pressure gradients to compete with the cosmic rays magnetic force at the scale $k_{\text{fast}}$ is simply $\beta_0\gtrsim 4$.
This condition is independent of the pressure anisotropy, which can be understood by noting that both the magnetic force and the pressure gradients force share the same (linear) dependency on the magnetic field.

The above estimates are valid if the anisotropy is exclusively due to the growth of the NR mode. However, pressure anisotropies with $P_m^\perp/P_m^{\smash[b]{\parallel}}>1$ can also drive the growth of the mirror mode. In the case when these two modes are competing (i.e. $\gamma_\text{mi}\sim \gamma$), the maximum level of anisotropy will then be determined by the mirror mode threshold pressure anisotropy: $A_\text{mi}=S_p/(\beta_m^\parallel-\beta_p)^{\alpha_p}$, where $\beta_m^\parallel=P_m^\parallel/W_B$ \cite{hellingerSolarWindProton2006}. The quantities $S_p$, $\beta_p$ and $\alpha_p$ are fitting parameters that are determined numerically for a given $\gamma_\text{mi}$. The rapid growth of the mirror mode also leads to important density fluctuations \cite{southwoodMirrorInstabilityPhysical1993} which would invalidate the incompressibility assumption in Eq. \ref{eq:anisotropic_evolution}, and thus make analytical estimates of $P_m^\perp$ and $P_m^\parallel$ near saturation unreliable, especially in environments with $\beta_0\gg 1$ where the mirror mode growth rate is maximum \cite{garyProtonTemperatureAnisotropy1997}. For these reasons, we restrict our analysis to regimes where $\beta_0\lesssim 1$, such that the parallel plasma beta in the amplified magnetic field is small $\beta_m^\parallel\ll 1$, and the mirror mode remains subdominant.

The parameters $\kappa^{-1}$ and $\beta_0$ can be used to assess the importance of Coulomb collisions and of pressure gradients in various environments. As an example, we consider the situation of a supernova shock propagating at a velocity $u_{cr}=5\times 10^3\ \text{km s}^{-1}$ in the interstellar medium with $n_m=1\ \text{cm}^{-3}$, $B=5\ \mu\text{G}$, $T_m=10^{4}\ \text{K}$, and a cosmic rays flux $n_{cr}u_{cr}=5\times 10^{4}\ \text{cm}^{-2}s^{-1}$ \cite{zweibelEnvironmentsMagneticField2010}. This gives $\beta_0=1.4$, $\gamma/\Omega_0=2.3\times 10^{-2}$, $A=33.0$ and a parameter $\kappa^{-1}= 6.8$, which is larger than the proton Coulomb collision frequency \cite{trubnikovParticleInteractionsFully1965} $\nu_0/\gamma_{max}= 6.2\times 10^{-3}$. Under such conditions, pressure anistropies should develop unimpeded and act to reduce the NR mode growth rate and saturated magnetic field.
In the context of laboratory experiments, and considering parameters typical of high power laser facilities: $u_{cr}=10^3\ \text{km s}^{-1}$, $n_m=10^{19}\ \text{cm}^{-3}$, $B=0.2\ \text{MG}$ and $T_m=10^6\ \text{K}$, and a proton flux $n_{cr}u_{cr}=10^{26}\ \text{cm}^{-2}s^{-1}$, one finds $\beta_0=8.7\times 10^{-1}$, $\gamma/\Omega_0=3.6\times 10^{-1}$, $A=3.3$ and $\kappa^{-1}=5.6$, smaller than the collision frequency $\nu_0/\gamma=29.7$. In this case pressure anisotropies should be mitigated, allowing for a faster growth of the magnetic perturbations.

\bibliography{main}

\end{document}